\documentclass[aps,prd,preprint,superscriptaddress,showpacs,showkeys]{revtex4}
\usepackage{graphicx}
\usepackage{amsfonts,amsmath,amssymb,bm}

\begin{document}


%
%

\title{Divergence-type theory of conformal fields}

\author{J. Peralta-Ramos}

\address{CONICET and Departamento de F\'isica, Facultad de Ciencias Exactas y Naturales, \\
Universidad de Buenos Aires-Ciudad Universitaria, Pabell\'on I, 1428 \\
Buenos Aires, Argentina\\
jperalta@df.uba.ar}

\author{E. Calzetta}
\address{CONICET and Departamento de F\'isica, Facultad de Ciencias Exactas y Naturales, \\
Universidad de Buenos Aires-Ciudad Universitaria, Pabell\'on I, 1428 \\
Buenos Aires, Argentina\\
calzetta@df.uba.ar}


\begin{abstract} 
We present a nonlinear hydrodynamical description of a conformal plasma within the framework of divergence-type theories (DTTs), which are not based on a gradient expansion. We compare the equations of the DTT and the second-order theory (based on conformal invariants), for the case of Bjorken flow. The approach to ideal hydrodynamics is faster in the DTT, indicating that our results can be useful in the study of early-time dynamics in relativistic heavy-ion collisions.
\end{abstract}

\maketitle

\section{Introduction}
Relativistic dissipative hydrodynamics has become a succesful tool to describe several stages of relativistic heavy-ion collisions (HICS) \cite{heinz09,rom09,houv}. The complete second-order stress-energy tensor of a strongly coupled conformal fluid, relevant to  Quark-Gluon Plasma dynamics, was given independently in Refs. \onlinecite{sonhydro} and \onlinecite{bat}, where it was shown that the hydrodynamic description of a conformal field does not belong to the conventional (or entropy-wise)  Israel-Stewart \cite{baier06,israel} formalism. 

In this paper we present a quadratic divergence-type theory (DTT) for a conformal field in flat space-time \cite{nos}. The DTT goes beyond second order theories based on conformal invariants, since it resums all-order veloctiy gradients, but is limited to the case where the second-order coefficients involving vorticity vanish. We compare numerical results for the DTT to those for the second-order theory, for Bjorken flow. The DTT approaches the ideal fluid behaviour faster than the second-order theory, thus being a suitable tool to describe early-time dynamics and shock-waves in heavy-ion collisions.  
 
\section{Conformal hydrodynamics as a DTT}

DTTs are interesting alternatives to the IS formalism, because they are more general and flexible \cite{geroch}. A DTT is fully characterized once the generating function $\chi$ and the source tensor $I^{\mu\nu}$ are given as ultra-local functions of the enlarged set of variables \footnote{We consider the case of vanishing chemical potential, suitable for a conformal fluid.} $\zeta^A=(\beta_\mu,\xi_{\mu\nu})$. Here, $\beta_\mu=u_\mu/T$ is the inverse temperature vector, with $u_\mu$ being the velocity vector, and $\xi_{\mu\nu}$ is a symmetric and traceless tensor characterizing departures from equilibrium (it vanishes in equilibrium). 
Introducing the symbol $A^\mu_B$ to denote the set $(T^{\mu\nu},A^{\mu\nu\rho})$, where $A^{\mu\nu\rho}$ is the flux tensor, and $I_B$ the set $(0,0,I_{\mu\nu})$, where $I_{\mu\nu}$ is the source tensor, the DTT is summed up in the equations
\begin{equation}
A^\mu_B = \frac{\partial \Phi^\mu}{\partial \zeta^B} ~,~~ S^\mu_{;\mu} = -I_B \zeta^B ~,~~ 
A^\mu_{B;\mu} = I_B ~~,
\end{equation}
where 
\begin{equation}
S^\mu = \Phi^\mu - \beta_\nu T^{\mu\nu} - A^{\mu\nu\rho} \xi_{\nu\rho}
\end{equation}
is the entropy current and $\Phi^\mu=\partial \chi/\partial \beta_\mu$ is the thermodynamic potential.

For a conformal theory, $T^{\mu\nu}$ is traceless \footnote{The Weyl anomaly is zero in flat space-time, which is the case considered here.} and $T^{\mu\nu}\rightarrow e^{(d+2)\omega}T^{\mu\nu}$ under a Weyl transformation $g_{\mu \nu}\rightarrow e^{-2\omega(x^\gamma)}g_{\mu\nu}$, where $\omega$ is a function of space-time coordinates $x^\gamma$. These two requirements allow us to find the most general generating function which is quadratic in $\xi^{\mu\nu}$, and to calculate \cite{nos} $T^{\mu\nu}$ and $A^{\mu\nu\rho}$. The linear part (in $\xi^{\mu\nu}$) of the source tensor is calculated requiring the DTT to reduce to Eckart's theory at first order in velocity gradients. The quadratic part is calculated requiring that the Second Law hold exactly (for arbitrary departures from equilibrium). Once we have $\chi$ and $I^{\mu\nu}$, the hydrodynamic equations follow. We refer the reader to Ref. \onlinecite{nos} for further details. The theory based on second-order conformal invariants \cite{sonhydro,bat} is recovered \footnote{Actually, we can recover the second-order theory only when the second-order transport coefficients $\lambda_{2,3}$ are zero, but we emphasize that shear-shear coupling (proportional to $\lambda_1$) is included. This does not represent a serious limitation to the application of the DTT to heavy-ion collisions.} by performing an expansion at second-order in velocity gradients. 

In the following, we will restrict ourselves to Bjorken flow, which represents a significant simplification of the equations obtained in Ref. \onlinecite{nos}. 

\section{Comparison between second-order theory and DTT for Bjorken flow}

Bjorken flow is a successful toy model of heavy-ion collisions in the mid-rapidity region, capturing the most important features of the hydrodynamical evolution. In Milne coordinates (proper time and rapidity)
$\tau=\sqrt{t^2-z^2}$ and $\psi=\textrm{arctanh}(z/t)$, the velocity is constant $(u^\tau , u^{\perp}, u^\psi) =(1, 0, 0)$, but the dynamics is non-trivial due to the fact that several Christoffel symbols are nonzero. 

The hydrodynamic equations of the second-order theory become ($p$ and $\rho=3p$ are the pressure and the energy density in the local rest frame)
\begin{equation}
\partial_\tau \rho = -\frac{\rho+p}{\tau}+\frac{\Pi^{\psi}_\psi}{\tau} ~~,~~
\partial_\tau \Pi^{\psi}_\psi = -\frac{\Pi^{\psi}_\psi}{\tau_\pi} +\frac{4\eta}{3\tau_\pi \tau} -\frac{4}{3\tau}\Pi^{\psi}_\psi - \frac{\lambda_1}{2\tau_\pi \eta^2}[\Pi^\psi_\psi]^2 ~,
\label{bjsec}
\end{equation}
where $\Pi^\psi_\psi$ is the dissipative part of the stress-energy tensor.

The equations of the DTT read
\begin{equation}
\partial_\tau \rho = -\frac{1}{\tau}\bigg(\rho+p -TF_1 [\xi^{\psi}_\psi]^2 \bigg) \\
+ \frac{\eta}{\tau}\xi^{\psi}_\psi
\label{bj1exactnew}
\end{equation}
with 
$F_1=2\lambda_1/(9\eta^2)$, and
\begin{equation}
\partial_\tau \xi^\psi_\psi = \frac{4\tau }{3T \tau_\pi} -\bigg( \frac{2}{3\tau}+\frac{\tau^2}{T\tau_\pi}\bigg)\xi^\psi_\psi -\frac{\lambda_1 T \tau^2}{3\tau_\pi \eta^5} [\xi^\psi_\psi]^2  ~.
\label{bj2exactnew}
\end{equation}

We will now present our numerical results for $\Pi^\psi_\psi$. In the DTT, this quantity is calculated as 
\begin{equation}
\Pi^\psi_\psi= \eta \xi_\psi^\psi +F_1 T[\xi^\psi_\psi]^2 ~.
\end{equation}
We focus on the strongly-coupled SYM plasma, for which we have \cite{sonhydro} 
$\tau_\pi = 2(2-\ln 2)\eta/(sT)$ and $\lambda_1=\eta/(2\pi T)$,
where $s$ is the entropy density.
In the following, we present the results for \footnote{This value is close to the upper bound set by comparing dissipative hydrodynamics to measurements. See, for instance, Ref. \onlinecite{rom09}.} $\eta/s=0.375$. As initial conditions, we set $\Pi^\psi_\psi(\tau_0)=0$, $\tau_0=0.5$ fm/c and $\rho(\tau_0)=10$ GeV/fm$^3$. 

\begin{figure}
\includegraphics{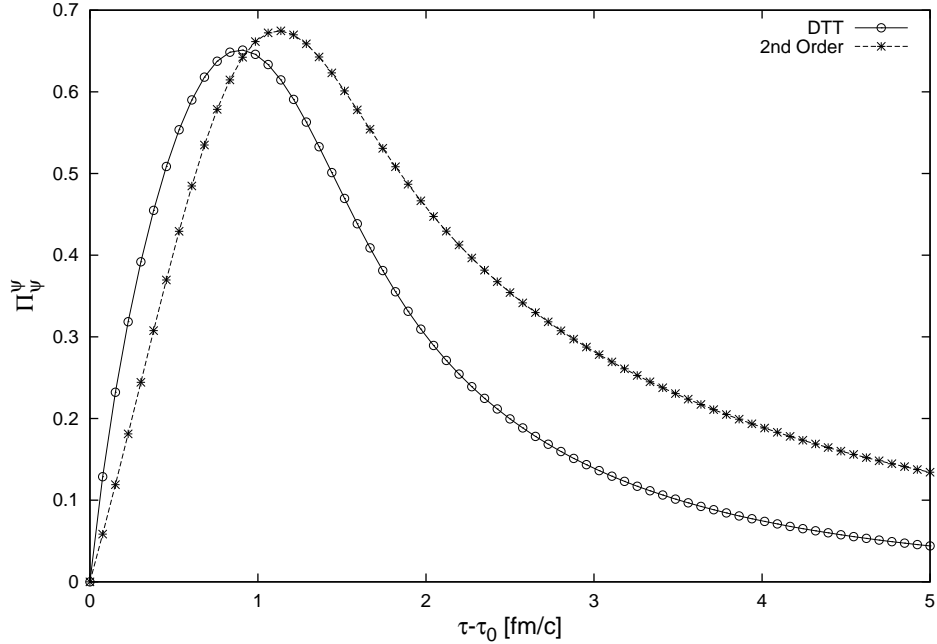}
\vspace{1cm}
\caption{Dissipative part of the stress-energy tensor as a function of proper time, for the DTT and the second-order theory with $\eta/s=0.375$.}
\label{f1}
\end{figure}

Fig. \ref{f1} shows the evolution of $\Pi^\psi_\psi$, for the DTT and the second-order theory.

With respect to the second-order theory, our results are in good agreement with those of previous studies \cite{houv,baier06}. From our results, we conclude that the relaxation towards ideal hydrodynamics is faster in the DTT than in the second-order theory. This means that, as expected on theoretical grounds, the hydrodynamic evolution in the DTT is closer to that obtained from transport theory.

\section{Conclusions}

We presented the hydrodynamic equations of the DTT for Bjorken flow, and compared them with those of the second-order theory. Our results show that the relaxation towards ideal hydrodyanamics is faster in the DTT as compared to the second-order theory. The DTT may be useful in the analysis of early-time dynamics and shock-waves in heavy-ion collisions, because it is not based on an expansion in velocity gradients. 

\section*{Acknowledgments}
We thank Paul Romatschke, Robert Geroch and Dirk Rischke for valuable comments and suggestions. This work has been supported in part by ANPCyT, CONICET and UBA (Argentina).



\begin{thebibliography}{00}    
\bibitem{heinz09} U. Heinz, arXiv:0901.4355 [nucl-th]; A. Muronga, Phys. Rev. C {\bfseries 69}, 034903 (2004); E. Shuryak, Prog. Part. Nucl. Phys. {\bfseries 62}, 48 (2009).
\bibitem{rom09} P. Romatschke, arXiv:0902.3663 [hep-ph].  
\bibitem{houv} P. Huovinen, and D. Molnar, arXiv:0808.0953 [nucl-th]. 
\bibitem{sonhydro} R. Baier, P. Romatschke, D. T. Son, A. O. Starinets, and M. A. Stephanov, J. High Energy Phys. {\bfseries 04}, 100 (2008).
\bibitem{bat} S. Bhattacharyya, V. E. Hubeny, S. Minwalla, and M. Rangamani, J. High Energy Phys. {\bfseries 02}, 45 (2008).
\bibitem{baier06} R. Baier, P. Romatschke, and U. A. Wiedemann, Phys. Rev. C {\bfseries 73}, 064903 (2006). 
\bibitem{israel} W. Israel, Ann. Phys. (NY) {\bfseries 100}, 310 (1976); W. Israel, and J. Stewart, Ann. Phys. (N.Y.) {\bfseries 118}, 341 (1979).
\bibitem{nos} J. Peralta-Ramos, and E. Calzetta, Phys. Rev. D {\bfseries 80}, 126002 (2009).
\bibitem{geroch} R. Geroch, and L. Lindblom, Phys. Rev. D {\bfseries 41}, 1855 (1990).

\end{thebibliography}
\end{document}